\documentclass[pteplogo]{ptephy}

\preprintnumber{XXXX-XXXX}

\usepackage{amsmath}
\usepackage{graphicx}
\usepackage{bm}

\newcommand{\dket}[1]{| #1\rangle\!\rangle}
\newcommand{\dbra}[1]{\langle\!\langle #1|}
\newcommand{\dbraket}[2]{\langle\!\langle #1| #2\rangle\!\rangle}
\newcommand{\bigpar}[1]{\biggl( #1 \biggr)}

\newcommand{\nbmat}[1]{{\bar\psi}^{(#1)}}
\newcommand{\ba}{a}
\newcommand{\bz}{{\bar z}}
\newcommand{\bt}{{\bar t}}
\newcommand{\h}[2]{\{#1_{#2}\}}

\begin{document}

\title{
Non-divergent Representation\\of Non-Hermitian Operator\\near the Exceptional Point\\with Application to a Quantum Lorentz Gas
}

\author{
\name{Kazunari Hashimoto}{1,\ast},
\name{Kazuki Kanki}{1},
\name{Hisao Hayakawa}{2},
and
\name{Tomio Petrosky}{3,4}}

\address{\affil{1}{Department of Physical Science, Osaka Prefecture University, Sakai 599-8531, Japan}
\affil{2}{Yukawa Institute for Theoretical Physics, Kyoto University, Kyoto 606-8501, Japan}
\affil{3}{Institute of Industrial Science, The University of Tokyo, Tokyo 153-8505, Japan}
\affil{4}{Center for Studies in Statistical Mechanics and Complex Systems, The University of Texas at Austin, TX 78712 USA}
\email{kazu.uncertainworld@gmail.com}}

\begin{abstract}
We propose a non-singular representation for a non-Hermitian operator even if the parameter space contains exceptional points (EPs), at which the operator cannot be diagonalized and the usual spectral representation ceases to exist.
Our representation has a generalized Jordan block form and is written in terms of extended pseudo-eigenstates.
Our method is free from a divergence in the spectral representation at EPs, at which multiple eigenvalues and eigenvectors coalesce and the eigenvectors cannot be normalized.
Our representation improves the accuracy of numerical calculations of physical quantities near EPs.
We also find that our method is applicable to various problems related to EPs in the parameter space of non-Hermitian operators.
We demonstrate the usefulness of our representation by investigating Boltzmann's collision operator in a one-dimensional quantum Lorentz gas in the weak coupling approximation.
\end{abstract}

\subjectindex{A52, A58, A64}

\maketitle

\section{Introduction}

The importance of non-Hermitian operator has been recognized in many areas of physics in recent years, both on an applied level and on a fundamental level.
Such non-Hermitian operators commonly appear, for example, when we discuss irreversible processes in open systems \cite{Balescu75, Resibois77, Heiss90, Heiss91, Heiss99, Heiss01}, as well as ${\cal PT}$ (parity-time) symmetric systems \cite{Klaiman08, Velle13, Candanedo14, Peng14}.
It is well known that an effective Liouvillian (which is called a collision operator in non-equilibrium statistical mechanics) for open quantum system takes a non-Hermitian form (see Appendix B and \cite{petpri97}).
An example of such non-Hermitian operator can be found when the partial trace of the environment is carried out \cite{Breuer02}.

Among many characteristic properties of the non-Hermitian operators, the appearance of exceptional points (EPs) in parameter space is especially interesting \cite{Kato} and it has been studied in many contexts in recent years \cite{Heiss90, Heiss91, Heiss99, Heiss01, Savannah12, Berry04, Heiss04, Cartarius07, Rubinstein07, Heiss08, Heiss10, Lefebvre09, Cartarius11, Demange12, Gilary13, Fuchs14, Dembowski01, Dembowski03, Dietz07, Lee09, Schindler11, Peng14}.
The EP is a singular point in the parameter space at which two eigenstates coalesce.
As a result the non-Hermitian operator is not diagonalizable at this point.
Instead, the operator can be reduced to the Jordan block form with the aid of a pseudo-eigenstate for a pair of coalesced eigenstates \cite{Demange12, Bhamathi}.
This singularity manifests itself in that the normalization constant of the eigenstate diverges at the EP \cite{Heiss01}.

Due to the divergence in the eigenstate representation, we encounter difficulties when we investigate physical properties of the system near the EPs.
They cause problems, for example, when we evaluate numerical values of physical quantities on the basis of the eigenstate expansion, because each contribution from the eigenstates becomes divergent.

In this paper, we propose a non-divergent representation of the non-Hermitian operator which is applicable in all regions in the parameter space, hence remove the difficulty mentioned above.
We obtain this representation by introducing a generalization of the pseudo-eigenstate that appears at the EPs for the Jordan block calculation.

In the present paper, we give the detailed description of the representation by a general $2\times 2$ non-Hermitian operator.
This is because the most likely case for an EP to occur is the two level coalescence \cite{Heiss08,Demange12} and all essential properties of the case can be illustrated with a $2\times 2$ matrix \cite{Heiss91,Heiss99}.

We apply this non-divergent representation to Boltzmann's collision operator for a one-dimensional (1D) perfect quantum Lorentz gas \cite{Balescu63, Zhang95} as a working example.
The Lorentz gas gives us one of the simplest examples of the non-Hermitian operators that has the EPs in the eigenvalue problem of the Liouvillian.
To our knowledge, this is the first time that the EP problem has been studied at the level of the Liouvillian.
We stress, however, that this representation is applicable to any non-Hermitian operators having EPs including the case of the effective Hamiltonian that appears in a wide class of quantum mechanics.

The structure of the paper is as follows:
In section 2 we summarize general properties of an EP with a general $2\times2$ non-Hermitian matrix.
In section 3 we derive a standard Jordan block structure at the EP in terms of the pseudo-eigenstate representation of the $2\times2$ matrix.
In section 4 we introduce an extended pseudo-eigenstate for the matrix to obtain a representation that remains non-divergent in all region of the parameter space.
In section 5 we present an application of our representation to the 1D perfect quantum Lorentz gas.
In the first three subsections of Sec.5 we give an introduction of the model and the derivation of Boltzmann's collision operator in terms of the theory of complex spectral representation of the Liouvillian \cite{petpri97}.
In subsection 5.4 we present spectrum and eigenstates of the collision operator.
There we also present standard Jordan block representation of the collision operator with the pseudo-eigenstates.
In subsection 5.5 we construct the extended pseudo-eigenstates for the system and present the extended Jordan block representation of the collision operator.
In subsection 5.6 we discuss the time evolution of the distribution function using both the eigenstates representation and the extended pseudo-eigenstates representation.
In subsection 5.7 we numerically estimate the time evolution of the system near the EP, and demonstrate that our non-divergent representation significantly improves the accuracy of the numerical calculation as compared with the calculation using the conventional eigenstate representation.
In subsection 5.8 we give some remarks on physical aspects of the EP in the spectrum of the Liouvillian and its relation to the transport process of the system.
In section 6 we summarize our results and give discussions.
In appendix A we discuss the generalization of the extended pseudo-eigenstate representation to the multiple coalescence of an arbitrary number $N$ of eigenstates at a so-called EP$N$ \cite{Heiss08}.
In appendix B we give a brief summary of the complex spectral representation of the Liouvillian for use in Sec.5.

\section{General Properties of an Exceptional Point}

Let us give a brief summary of mathematical properties of EPs.
Mathematically, the {\it exceptional points} (EPs) are defined as branch point singularities in the parameter space of the spectrum of a matrix at which two or more eigenstates coincide \cite{Kato}.
Its significant difference from the usual degeneracy lies in the fact that both the eigenvalues and the corresponding eigenstates coalesce.
Since the Hermitian operator has a complete set of eigenstates, EPs occur only in the spectrum of non-Hermitian operator.

Over the wide range of non-Hermitian physical systems, the most common case for an EP to occur is the two level coalescences since multiple level coalescence requires that more parameters should be fine-tuned to satisfy conditions for coalescence \cite{Heiss08,Demange12}.
Therefore, in the present paper, we restrict our detailed study to the two level coalescence.
Generalization of our main result to the multiple coalescence is straightforward and is briefly presented in Appendix A.
In the case of a two level coalescence, we can illustrate all essential properties of the EPs with a $2\times2$ matrix.
Indeed, for both finite and infinite dimensional matrices, an isolated EP can be effectively described by employing a $2\times2$ matrix \cite{Heiss91,Heiss99}.
Here, we consider a general $2\times 2$ non-Hermitian matrix in the following form,
	\begin{equation}
	{\hat M}=
	\begin{pmatrix}
	a & b \\
	c & d
	\end{pmatrix},
	\label{eq:general-model}
	\end{equation}
where $a$, $b$, $c$ and $d$ are complex parameters.
Let us now denote its eigenvalue as $z_{\alpha}$ and right- or left-eigenvectors as ${\bf u}_{\alpha}$ or ${\tilde{\bf u}}_{\alpha}^{\dagger}$, i.e.,
	\begin{equation}
	{\hat M}{\bf u}_{\alpha}=z_{\alpha}{\bf u}_{\alpha},\hspace{10pt}{\tilde{\bf u}}_{\alpha}^{\dagger}{\hat M}=z_{\pm}{\tilde{\bf u}}_{\alpha}^{\dagger}.
	\label{eq:general-eigeneq}
	\end{equation}
The eigenvalues of Eqs.(\ref{eq:general-model}) and (\ref{eq:general-eigeneq}) are given by
	\begin{equation}
	z_{\pm}=\frac{a+d}{2}\pm\frac{1}{2}[(a-d)^{2}+4bc]^{1/2}.
	\label{eq:general-eigenvalues}
	\end{equation}
The corresponding right- and left-eigenstates are given by
	\begin{subequations}
	\begin{equation}
	{\bf u}_{\pm}=
	\begin{pmatrix}
	\biggr[1\pm\frac{[(a-d)^{2}+4bc]^{1/2}}{a-d}\biggr]^{1/2} \\
	-\frac{(-bc)^{1/2}}{b}\frac{a-d}{[(a-d)^{2}]^{1/2}}\biggr[1\mp\frac{[(a-d)^{2}+4bc]^{1/2}}{a-d}\biggr]^{1/2}
	\end{pmatrix},
	\end{equation}
and
	\begin{equation}
	{\tilde{\bf u}}_{\pm}^{\dagger}\equiv\Biggr((-bc)^{1/2}\frac{a-d}{[(a-d)^{2}]^{1/2}}\biggr[1\pm\frac{[(a-d)^{2}+4bc]^{1/2}}{a-d}\biggr]^{1/2},b\biggr[1\mp\frac{[(a-d)^{2}+4bc]^{1/2}}{a-d}\biggr]^{1/2}\Biggr).
	\end{equation}
	\label{eq:general-eigenvectors}
	\end{subequations}
Here, we have not normalized the eigenstates (\ref{eq:general-eigenvectors}) considering the fact that they cannot be normalized at the EPs.
The inner products of these right- and left-eigenstates are given by
	\begin{equation}
	({\tilde{\bf u}}_{\pm},{\bf u}_{\pm})\equiv{\tilde{\bf u}}^{\dagger}_{\pm}\cdot{\bf u}_{\pm}=\pm2(-bc)^{1/2}\frac{[(a-d)^{2}+4bc]^{1/2}}{[(a-d)^{2}]^{1/2}}.
	\label{eq:general-inner}
	\end{equation}
Here we define the square root of a complex number
	\begin{equation}
	z=|z|e^{i\theta},\hspace{10pt}(-\pi<\theta\leq\pi)
	\end{equation}
as
	\begin{equation}
	z^{1/2}=\sqrt{|z|}e^{i\theta/2}.
	\end{equation}

At the points in the parameter space satisfying
	\begin{equation}
	(a-d)^{2}+4bc=0,
	\end{equation}
both eigenvalues (\ref{eq:general-eigenvalues}) and eigenvectors (\ref{eq:general-eigenvectors}) have branch point singularities in the parameter space, namely the EPs.
At these points, the eigenvalues degenerate as
	\begin{equation}
	z_{\pm}=z_{{\rm EP}}\equiv\frac{a+d}{2}.
	\end{equation}
Moreover, at these points, the eigenstates (\ref{eq:general-eigenvectors}) also ``degenerate'' in the sense that these two eigenstates collapse into a single eigenstates as
	\begin{subequations}
	\begin{equation}
	{\bf u}_{\pm}={\bf u}_{{\rm EP}}\equiv
	\begin{pmatrix}
	1 \\
	-\frac{(-bc)^{1/2}}{b}\frac{a-d}{[(a-d)^{2}]^{1/2}}
	\end{pmatrix},
	\end{equation}
and
	\begin{equation}
	{\tilde{\bf u}}_{\pm}^{\dagger}={\tilde{\bf u}}_{{\rm EP}}^{\dagger}=b\Biggr(\frac{(-bc)^{1/2}}{b}\frac{a-d}{[(a-d)^{2}]^{1/2}},\;1\Biggr).
	\end{equation}
	\end{subequations}
Since there is only one linearly independent eigenstates at the EPs, the operator (\ref{eq:general-model}) is non-diagonalizable at these points.
This collapse of eigenstates does not take place at the usual degeneracy point in a Hermitian operator, where a degenerate eigenvalue is shared by two distinct eigenstates \cite{Heiss01}.
For this reason, such a degeneracy point in the present case, namely EP, is often called a {\it non-Hermitian degeneracy point} \cite{Berry04} to emphasize that it appears only in non-Hermitian operators.

Note that the norm of eigenstates vanishes at the EPs (see Eq.(\ref{eq:general-inner})).
Except at the EPs, the eigenstates (\ref{eq:general-eigenvectors}) are normalized as
	\begin{subequations}
	\begin{equation}
	{\bf v}_{\pm}\equiv{\bf u}_{\pm}\bigr/[({\tilde{\bf u}}_{\pm},{\bf u}_{\pm})]^{1/2},
	\end{equation}
	\begin{equation}
	{\tilde{\bf v}}_{\pm}^{\dagger}\equiv{\tilde{\bf u}}_{\pm}^{\dagger}\bigr/[({\tilde{\bf u}}_{\pm},{\bf u}_{\pm})]^{1/2}
	\end{equation}
	\label{eq:general-normal-eigenvectors}
	\end{subequations}
Then, they satisfy the following bi-orthonormality and bi-completeness relations,
	\begin{equation}
	({\tilde{\bf v}}_{\alpha},{\bf v}_{\alpha'})=\delta_{\alpha,\alpha'},
	\end{equation}
	\begin{equation}
	\sum_{\alpha}{\bf v}_{\alpha}{\tilde{\bf v}}_{\alpha}^{\dagger}={\hat I},
	\end{equation}
where $\alpha$ and $\alpha'$ take the values ``$+$'', ``$-$'' and ${\hat I}$ is the unit matrix of size $2$.

\section{Jordan Block Representation at the EP and the Pseudo-eigenstate}

In this section, we summarize the well known Jordan block structure at the EP and its relation to the pseudo-eigenstate \cite{Demange12,Bhamathi} in order to prepare for the introduction of the extended pseudo-eigenstate in the next section.
The right pseudo-eigenstate, denoted by ${\bf u}'_{{\rm EP}}$, is defined through the following relation,
	\begin{equation}
	[{\hat M}-z_{{\rm EP}}{\hat I}]{\bf u}'_{{\rm EP}}={\bf u}_{{\rm EP}},
	\label{eq:general-Jordanchain-right}
	\end{equation}
which is called the Jordan chain relation \cite{Demange12}.
Similarly, the left pseudo-eigenstate ${\tilde{\bf u}}_{{\rm EP}}^{\prime\;\dagger}$ is defined by
	\begin{equation}
	{\tilde{\bf u}}_{{\rm EP}}^{\prime\;\dagger}[{\hat M}-z_{{\rm EP}}{\hat I}]={\tilde{\bf u}}_{{\rm EP}}^{\dagger}.
	\label{eq:general-Jordanchain-left}
	\end{equation}
We then have
	\begin{equation}
	{\bf u}'_{{\rm EP}}=\frac{1}{2}
	\begin{pmatrix}
	\frac{1}{(-bc)^{1/2}}\frac{[(a-d)^{2}]^{1/2}}{a-d} \\
	\frac{1}{b}
	\end{pmatrix},\hspace{10pt}
	{\tilde{\bf u}}_{{\rm EP}}^{\prime\;\dagger}=\frac{1}{2}\Biggr(1,\;\frac{(-bc)^{1/2}}{c}\frac{a-d}{[(a-d)^{2}]^{1/2}}\Biggr).
	\label{eq:general-pseudo}
	\end{equation}
These pseudo-eigenstates satisfy the following bi-orthonormality relations
	\begin{equation}
	({\tilde{\bf u}}_{{\rm EP}},{\bf u}'_{{\rm EP}})=1,\hspace{5pt}({\tilde{\bf u}}_{{\rm EP}}^{\prime},{\bf u}_{{\rm EP}})=1,\hspace{5pt}({\tilde{\bf u}}_{{\rm EP}},{\bf u}_{{\rm EP}})=0,\hspace{5pt}({\tilde{\bf u}}_{{\rm EP}}^{\prime},{\bf u}'_{{\rm EP}})=0,
	\end{equation}
and the bi-completeness relation
	\begin{equation}
	{\bf u}_{{\rm EP}}{\tilde{\bf u}}_{{\rm EP}}^{\prime\;\dagger}+{\bf u}'_{{\rm EP}}{\tilde{\bf u}}_{{\rm EP}}^{\dagger}={\hat I}.
	\end{equation}
In term of this basis, the operator ${\hat M}$ is represented by the standard Jordan block,
	\begin{equation}
	{\hat M}=
	\bordermatrix{
	& {\bf u}_{{\rm EP}} & {\bf u}'_{{\rm EP}} \cr
	{\tilde{\bf u}}_{{\rm EP}}^{\prime\;\dagger} & z_{{\rm EP}} & 1 \cr
	{\tilde{\bf u}}_{{\rm EP}}^{\dagger} & 0 & z_{{\rm EP}} \cr
	}.
	\label{eq:general-jordan-matrix}
	\end{equation}
As a result, the operator ${\hat M}$ is represented by the Jordan block form by introducing the pseudo-eigenstate at the EPs.
However, the representation does not remove the divergent behavior of eigenstates (\ref{eq:general-normal-eigenvectors}) near the EPs since it is only applicable at the EPs.

\section{The Extended Pseudo-eigenstate Representation}

So far, we have shown that, when the operator ${\hat M}$ has EPs, it is not diagonalizable at these points.
Instead, the operator can be reduced to a Jordan block form at these points.
In this section, we introduce a representation which does not have any singularity at EPs by extending the concept of the pseudo-eigenstate representation to a non-exceptional point.

At a non-exceptional point, there exist two linearly independent eigenstates ${\bf u}_{+}$ and ${\bf u}_{-}$.
For ${\bf u}_{+}$, for example, we introduce a {\it right extended pseudo-eigenstate} for $z_{-}$, denoted by ${\bf u}'_{-}$, which satisfies the following {\it extended Jordan chain relation} for an arbitrary point in the parameter space,
	\begin{equation}
	[{\hat M}-z_{-}{\hat I}]{\bf u}'_{-}={\bf u}_{+}.
	\label{eq:general-extended-Jordanchain-right-1}
	\end{equation}
On the other hand, for ${\bf u}_{+}$, we introduce a right extended pseudo-eigenstate ${\bf u}'_{+}$
	\begin{equation}
	[{\hat M}-z_{+}{\hat I}]{\bf u}'_{+}={\bf u}_{-}.
	\label{eq:general-extended-Jordanchain-right-2}
	\end{equation}
Similarly, we can introduce a {\it left extended pseudo-eigenstate} for $z_{\pm}$, denoted by ${\tilde{\bf u}}_{\pm}^{\prime\;\dagger}$, satisfying
	\begin{equation}
	{\tilde{\bf u}}_{\pm}^{\prime\;\dagger}[{\hat M}-z_{\pm}{\hat I}]={\tilde{\bf u}}_{\mp}^{\dagger}.
	\label{eq:general-extended-Jordanchain-left}
	\end{equation}
We impose the normalization conditions for the right and left extended pseudo-eigenstates,
	\begin{equation}
	({\tilde{\bf u}}_{\pm},{\bf u}'_{\pm})=1,\hspace{10pt}({\tilde{\bf u}}_{\pm}^{\prime},{\bf u}_{\pm})=1.
	\label{eq:normal-extended-jordan-basis}
	\end{equation}
Then, we have right and left pseudo-eigenstates
	\begin{subequations}
	\begin{equation}
	{\bf u}'_{\pm}=\frac{1}{2}
	\begin{pmatrix}
	\frac{1}{(-bc)^{1/2}}\frac{[(a-d)^{2}]^{1/2}}{a-d}\biggr[1\pm\frac{[(a-d)^{2}+4bc]^{1/2}}{a-d}\biggr]\\
	\frac{1}{b}\biggr[1\mp\frac{[(a-d)^{2}+4bc]^{1/2}}{a-d}\biggr]
	\end{pmatrix}.
	\end{equation}
and
	\begin{equation}
	{\tilde{\bf u}}_{\pm}^{\prime\;\dagger}=\frac{1}{2}\Biggr(\biggr[1\pm\frac{[(a-d)^{2}+4bc]^{1/2}}{a-d}\biggr],\;\frac{(-bc)^{1/2}}{c}\frac{a-d}{[(a-d)^{2}]^{1/2}}\biggr[1\mp\frac{[(a-d)^{2}+4bc]^{1/2}}{a-d}\biggr]\Biggr),
	\end{equation}
	\label{eq:general-extended-pseudo}
	\end{subequations}
respectively.
They satisfy the bi-orthogonality relation
	\begin{equation}
	({\tilde{\bf u}}_{\pm}^{\prime},{\bf u}'_{\mp})=0.
	\label{eq:general-orthogonal-extended-jordan-basis}
	\end{equation}
The two sets of vectors $\{{\bf u}_{+},{\bf u}'_{-},{\tilde{\bf u}}_{+}^{\prime\;\dagger},{\tilde{\bf u}}_{-}^{\dagger}\}$ and $\{{\bf u}_{-},{\bf u}'_{+},{\tilde{\bf u}}_{-}^{\prime\;\dagger},{\tilde{\bf u}}_{+}^{\dagger}\}$, respectively, form a bi-complete basis,
	\begin{equation}
	{\bf u}_{\pm}{\tilde{\bf u}}_{\pm}^{\prime\;\dagger}+{\bf u}'_{\mp}{\tilde{\bf u}}_{\mp}^{\dagger}={\hat I}.
	\end{equation}
In terms of either basis set, the collision operator is represented by the Jordan block-like matrix for an arbitrary point in the parameter space as
	\begin{equation}
	{\hat M}=
	\bordermatrix{
	& {\bf u}_{\pm} & {\bf u}'_{\mp} \cr
	{\tilde{\bf u}}_{\pm}^{\prime\;\dagger} & z_{\pm} & 1 \cr
	{\tilde{\bf u}}_{\mp}^{\dagger} & 0 & z_{\mp} \cr
	}
	\label{eq:general-extend-jordan}
	\end{equation}
Note that this matrix differs from the Jordan block (\ref{eq:general-jordan-matrix}) because both eigenvalues $z_{+}$ and $z_{-}$ appear on the diagonal.
By taking a limit to an EP for (\ref{eq:general-extend-jordan}), we recover the Jordan block representation (\ref{eq:general-jordan-matrix}) just at the EP.
The extended pseudo-eigenstates (\ref{eq:general-extended-pseudo}) also reduces to the usual pseudo-eigenstates (\ref{eq:general-pseudo}) in the limit.

Here, we have introduced the extended pseudo-eigenstate representation for a general $2\times2$ non-Hermitian operators.
Its generalization to a situation where multiple coalescence occurs at an EP is obvious and we will present the generalization in Appendix A.

\section{Application to Weakly Coupled One-dimensional Quantum Perfect Lorentz Gas}

As an illustration of the representation introduced in the previous section, here we present a physical example.
As a woking example, we consider Boltzmann's collision operator for a weakly-coupled one-dimensional (1D) quantum perfect Lorentz gas \cite{Balescu63,Zhang95}.
We will show that our representation provides a numerically stable representation near the EPs and it will significantly improve accuracy of numerical estimation of the time evolution of the system near the EPs as compared with the conventional eigenstate representation (see subsections 5.6 and 5.7).

The Lorentz gas is one of the simplest example of the non-Hermitian operator that has the EPs in the eigenvalue problem in the Liouville space.
Since almost all examples discussed in the literature on the problem on EPs are on the problem of the Hamiltonian or some phenomenological equation of motion in the ${\cal PT}$-symmetric systems, it is worthwhile to add an example in the Liouville space description, in particular, in the kinetic theory in irreversible statistical mechanics.

In the first three subsections, we shall give derivation of Boltzmann's collision operator for the system and show that it reduces to a $2\times2$ non-Hermitian matrix of the form (\ref{eq:general-model}) in the Wigner representation as shown in the expression (\ref{eq:nmatbcoll}) in the subsection 5.3.
The reader who is not interested in the derivation can skip these subsections and directly go to the subsection $5.4$.

\subsection{The weakly-coupled 1D quantum perfect Lorentz gas}

The Lorentz gas consists of one light-mass particle (the test particle) with mass $m$ and $N$ heavy particles with mass $M$.
The Hamiltonian of the system is given by
	\begin{equation}
	H=H_{0}+gV=\frac{p^{2}}{2m}+\sum_{j=1}^{N}\frac{p_{j}^{2}}{2M}+g\sum_{j=1}^{N}V(|x-x_{j}|),
	\label{Hamiltonian}
	\end{equation}
where $g$ is the coupling constant and the interaction is assumed to be a short-range repulsive force.
In this paper, we shall consider the weak-coupling regime ($g\ll1$).
We also restrict our interest to the case $m/M\rightarrow0$, which is called the {\it perfect Lorentz gas} \cite{Balescu63}.
We suppose that the system is enclosed in a large 1D box of volume $L$ with the periodic boundary condition.
Hence, the interaction potential is expanded in the Fourier series as
	\begin{equation}
	V(|x-x_{j}|)=\frac{1}{L}\sum_{q}V_{|q|}e^{iq(x-x_{j})},
	\label{eq:potential-exp}
	\end{equation}
where $q$'s are integer multiples of $2\pi/L$.

In this paper, we shall consider the thermodynamic limit,
	\begin{equation}
	L\rightarrow\infty,\hspace{10pt}N\rightarrow\infty,\hspace{10pt}n\equiv\frac{N}{L}={\rm finite},
	\end{equation}
where $n$ is the concentration of heavy particles.
In this limit, the wavenumber and the momentum become continuous variables.
Hence we shall replace a summation of an integration and a Kronecker delta $\delta^{Kr}$ with a Dirac $\delta$-function as
	\begin{equation}
	\frac{2\pi}{L}\sum_{q}\rightarrow\int dq,\hspace{10pt}\frac{L}{2\pi\hbar}\delta^{Kr}(P-P')\rightarrow\delta(P-P'),
	\end{equation}
at an appropriate stage.

In this paper we investigate the time evolution of the reduced density matrix for the test particle, which is defined as
	\begin{equation}
	f(t)\equiv{\rm Tr}_{{\rm hev}.}[\rho(t)],
	\label{eq:redmat}
	\end{equation}
where ${\rm Tr}_{{\rm hev}.}$ denotes a partial trace over the heavy particles.
This procedure is equivalent to that used for quantum master equations for open quantum systems. \cite{Breuer02}

We assume that the initial condition of the system is given by
	\begin{equation}
	\rho(0)=f(0)\otimes\rho^{eq}_{{\rm hev}.},
	\end{equation}
where $\rho^{eq}_{{\rm hev}.}$ is the Maxwell distribution of the heavy particles with temperature $T$,
	\begin{equation}
	\rho^{eq}_{{\rm hev}.}=\prod^{N}_{j=1}\frac{\exp(-p_{j}^{2}/2Mk_{B}T)}{{\rm Tr}[\exp(-p_{j}^{2}/2Mk_{B}T)]},
	\end{equation}
where $k_{B}$ is the Boltzmann constant.
In the thermodynamic limit the time evolution of the density matrix associated with the heavy particles is negligible since its deviation from $\rho^{eq}_{{\rm hev}.}$ is proportional to $1/L$ in this limit, as can be easily shown.

\subsection{The Liouville space description}

The time evolution of the system is governed by the Liouville-von Neumann equation for the density matrix $\rho(t)$,
	\begin{equation}
	i\frac{\partial}{\partial t}\rho(t)=L_{H}\rho(t).
	\label{eq:liouville-eq}
	\end{equation}
Here $L_{H}$ is the Liouville-von Neumann operator (Liouvillan in short) which is defined by the commutation relation with the Hamiltonian of the system (\ref{Hamiltonian}),
	\begin{equation}
	L_{H}\rho\equiv\frac{1}{\hbar}[H,\rho].
	\end{equation}

To discuss the space and momentum dependence of the distribution of the particles in parallel with classical mechanics, it is convenient to introduce the Wigner distribution function:
	\begin{equation}
	\rho^{W}(X,\h{X}{j},P,\h{P}{j},t)\equiv\frac{1}{L^{N+1}}\sum_{k,\h{k}{j}}\rho_{k,\h{k}{j}}(P,\h{P}{j},t)e^{i(kX+k_{1}X_{1}+\cdots+k_{N}X_{N})},
	\label{eq:10}
	\end{equation}
which is a quantum analog of the phase space distribution function \cite{petpri97}.
Here the notation $\h{X}{j}$ represents a set of variables for the $N$ heavy particles and
	\begin{eqnarray}
	\rho_{k,\h{k}{j}}(P,\h{P}{j},t)&\equiv&\biggr\langle P+\frac{\hbar}{2}k,\biggr\{P_{j}+\frac{\hbar}{2}k_{j}\biggr\}\biggr|\rho(t)\biggr|P-\frac{\hbar}{2}k,\biggr\{P_{j}-\frac{\hbar}{2}k_{j}\biggr\}\biggr\rangle\nonumber\\
	&\equiv&\dbraket{k,\h{k}{j};P,\h{P}{i}}{\rho(t)},
	\label{def:wignerrepresentation}
\end{eqnarray}
where the single bra-ket vectors stand for vectors in the wave function space and the double bra-ket vectors stand for vectors in the Liouville space \cite{petpri97}.
Here the ``wavenumbers'' and the ``momenta'' in the Wigner representation are defined as
	\begin{equation}
	k\equiv\frac{p-p'}{\hbar},\hspace{10pt} P\equiv\frac{p+p'}{2},
	\end{equation}
and the Wigner basis is defined by a dyad of two eigenstates of $H_{0}$ as
	\begin{equation}
	\dket{k,\h{k}{j};P,\h{P}{j}}\equiv|p,\h{p}{j}\rangle\!\langle p',\h{p'}{j}|.
	\end{equation}
We represent a linear operator $A$ in the wave function space as a ket-vector $\dket{A}$ in the Liouville space.
The inner product of the bra- and ket-vectors is then defined by
	\begin{equation}
	\dbraket{B}{A}={\rm Tr}[B^{\dagger}A],
	\end{equation}
where $B^{\dagger}$ is the Hermitian conjugate of a liner operator $B$.
As a result, it is easy to show that the Wigner basis vectors are normalized with respect to the box normalization condition
	\begin{equation}
	\dbraket{k,k,\!\h{k}{j};\!P,\!\h{P}{j}}{k',\!\h{k'}{j};\!P',\!\h{P'}{j}}\!=\!\delta^{Kr}(k-k')\delta^{Kr}(P-P')\!\prod_{j=1}^{N}\!\delta^{Kr}(k_{j}-k'_{j})\delta^{Kr}(P_{j}-P'_{j}).
	\end{equation}

\subsection{Derivation of Boltzmann's collision operator for the 1D quantum perfect Lorentz gas}

To obtain time evolution equation for the reduced density matrix for the test particle from the Liouville equation, we consider eigenvalue problem of the Liouvillian by applying the theory of the complex spectral representation of the Liouvillian.
Its general formalism is briefly summarized in appendix B.

In the Liouville space, the eigenvalue problem of the Liouvillian is given by
	\begin{equation}
	L_{H}\dket{F^{(k)}_{\alpha}}=Z^{(k)}_{\alpha}\dket{F^{(k)}_{\alpha}},\hspace{10pt}\dbra{{\tilde F}^{(k)}_{\alpha}}L_{H}=Z^{(k)}_{\alpha}\dbra{{\tilde F}^{(k)}_{\alpha}},
	\label{eq:lorentz-eigenequation}
	\end{equation}
where we have denoted the right-eigenstate as $\dket{F^{(k)}_{\alpha}}$ and left-eigenstate as $\dbra{{\tilde F}^{(k)}_{\alpha}}$.
Here, the index $\alpha$ specifies the eigenstate in the correlation subspace associated with a projection operator $P^{(k)}$, which acts on the distribution function (\ref{eq:10}) as
	\begin{equation}
	P^{(k)}\rho^{W}(X,\h{X}{j},P,\h{P}{j},t)=\rho_{k,\h{0}{}}(P,\h{P}{j},t),
	\label{eq:16}
	\end{equation}
and it satisfies
	\begin{equation}
	gP^{(k)}L_{V}P^{(k)}=0.
	\label{eq:potential-condition}
	\end{equation}
Using the double bra-ket notation, the projection operator is written as
	\begin{equation}
	P^{(k)}\equiv\sum_{P}\sum_{\{P_{j}\}}\dket{k,\{0\};P,\{P_{j}\}}\!\dbra{k,\{0\};P,\{P_{j}\}}.
	\end{equation}
We also define its complement as $Q^{(k)}\equiv1-P^{(k)}$.

By applying these projection operators on Eqs.(\ref{eq:lorentz-eigenequation}), the eigenvalue equation of the Liouvillian takes the form
	\begin{equation}
	\Psi^{(k)}(Z^{(k)}_{\alpha})P^{(k)}\dket{F^{(k)}_{\alpha}}=Z^{(k)}_{\alpha}P^{(k)}\dket{F^{(k)}_{\alpha}},
	\end{equation}
where $\Psi^{(k)}(z)$ is the effective Liouvillian (\ref{eq:collop}) that is called the collision operator.
In the weak-coupling situation, the collision operator can be approximated up to the second order in $g$ as
	\begin{equation}
	\Psi^{(k)}_{2}(z)=P^{(k)}L_{0}P^{(k)}+g^{2}P^{(k)}L_{V}Q^{(k)}\frac{1}{z-L_{0}}Q^{(k)}L_{V}P^{(k)},
	\label{eq:21}
	\end{equation}
with
	\begin{equation}
	P^{(k)}L_{0}P^{(k)}\dket{k,\{0\};P,\{P_{j}\}}=\frac{kP}{m}\dket{k,\{0\};P,\{P_{j}\}}.
	\end{equation}
Note that, in the expression (\ref{eq:21}), the first-order term in $g$ vanishes according to the condition (\ref{eq:potential-condition}).

In this paper, we shall study a situation where the wavenumber $k$ satisfies
	\begin{equation}
	|k|\lesssim k_{P},
	\label{eq:boltzmann-region}
	\end{equation}
where
	\begin{equation}
	k_{P}=\frac{1}{l_{P}}\equiv\frac{\gamma_{P}}{2|P|/m},\hspace{10pt}\gamma_{P}\equiv g^{2}\frac{8\pi^{2}mn}{\hbar^{2}|P|}\bigr|V_{\frac{2P}{\hbar}}\bigr|^{2}.
	\label{def:k_P}
	\end{equation}
Here, $\gamma_{P}$ is the momentum relaxation rate of the test particle, which is evaluated using Fermi's golden rule.
We consider the case where the interaction range which we denote $d$ is much shorter than the mean-free-length $l_P$,
	\begin{equation}
	d\ll l_{P}.
	\end{equation}
Hence, a typical value of $q$ appearing in (\ref{eq:potential-exp}) is much larger than $k$ in (\ref{eq:21}),
	\begin{equation}
	|k|\ll |q|.
	\label{eq:k-smaller-than-l}
	\end{equation}
For this case with the weak coupling $g\ll1$, we can approximate the collision operator as
	\begin{equation}
	\Psi^{(k)}_{2}(Z^{(k)}_{\alpha})=\Psi^{(k)}_{2}(+i0)+O(g^{4}),
	\label{eq:bol1}
	\end{equation}
where $+i0$ means that the collision operator $\Psi^{(k)}_{2}(z)$ is evaluated on the real axis approaching from the upper half-plane to ensure the time evolution is properly oriented to the future $t>0$ \cite{petpri97}.

Let us now introduce the reduced collision operator acting on the reduced density matrix of the test particle (\ref{eq:redmat}) as
	\begin{equation}
	\psi^{(k)}\equiv{\rm Tr}_{{\rm hev.}}\Bigr[\Psi^{(k)}_{2}(+i0)\rho^{eq}_{{\rm hev.}}\Bigr].
	\label{eq:redcollop}
	\end{equation}
In the thermodynamic limit, the matrix element of this operator in the Wigner representation is given by
	\begin{equation}
	(\!(k,P|\psi^{(k)}|k,P')\!)=\biggr[\frac{kP}{m}-\frac{2\pi g^{2}n}{\hbar^{2}}\lim_{\epsilon\rightarrow+0}\!\int^{\infty}_{-\infty}\!\!\!dq|V_{|q|}|^{2}\partial^{\hbar q/2}_{P}\frac{1}{+i\epsilon-qP/m}\partial^{\hbar q/2}_{P}\biggr]\delta(P-P'),
	\label{eq:collop-matel}
	\end{equation}
with the reduced state of the test particle
	\begin{equation}
	|k;P)\!)\equiv\frac{L}{2\pi\hbar}\dket{k;P},
	\end{equation}
that is normalized by the $\delta$-function in the continuous spectrum limit.
Here, the operator $\partial^{\hbar q/2}_{P}$ is a displacement operator defined by
	\begin{equation}
	\partial^{\hbar q/2}_{P}\equiv e^{\frac{\hbar}{2}q\frac{\partial}{\partial P}}-e^{-\frac{\hbar}{2}q\frac{\partial}{\partial P}},
	\end{equation}
where $\exp{[a\partial/\partial P]}$ acts on a function of $P$ as $\exp{[a\partial/\partial P]}f(P)=f(P+a)$.
Furthermore, we have ignored the $k$ which appears in the denominator in (\ref{eq:collop-matel}) as compared with $q$ (see (\ref{eq:k-smaller-than-l})).
Note that the expression (\ref{eq:collop-matel}) does not depend on the temperature of the heavy particles $T$.
This is because in the limit of the perfect Lorentz gas $m/M\rightarrow0$ there is no energy transfer between the test particle and the heavy particles.

Performing the $q$ integration in (\ref{eq:collop-matel}), a matrix element of the collision operator $\psi^{(k)}$ is expressed as
	\begin{equation}
	(\!(k;P|\psi^{(k)}|k;P')\!)=\biggr(\frac{kP}{m}-i\frac{\gamma_{P}}{2}\biggr)\delta(P-P')+i\frac{\gamma_{P}}{2}\delta(P+P').
	\end{equation}
Hence, it has non-vanishing matrix elements only between the states $|k;P)\!)$ and $|k;-P)\!)$.
Physically, this is because there are only forward and backward scattering in this 1D system.
Therefore, in terms of this basis, the collision operator is represented by the $2\times2$ matrix
	\begin{equation}
	{\bar\psi}^{(a)}\equiv\frac{2}{\gamma_{P}}\psi^{(k)},
	\end{equation}
where
	\begin{equation}
	{\bar\psi}^{(a)}=\Bigr({\bar\psi}^{(a)}_{\mu\mu'}\Bigr)\equiv
	\begin{pmatrix}
	a & i\\
	i & -a
	\end{pmatrix}
	-i{\hat I},
	\label{eq:nmatbcoll}
	\end{equation}
where $\mu$ and $\mu'$ take values $P$ or $-P$, $a$ is the non-dimensionalized wavenumber defined by 
	\begin{equation}
	a\equiv\frac{k}{k_{P}},
	\end{equation}
and ${\hat I}$ is the unit matrix of size $2$.

In terms of the collision operator, the time evolution equation for the reduced density matrix for the test particle is given by
	\begin{equation}
	i\frac{\partial}{\partial t}{\hat p}^{(k)}\dket{f(t)}=\psi^{(k)}{\hat p}^{(k)}\dket{f(t)},
	\label{eq:kineticeq}
	\end{equation}
where
	\begin{equation}
	{\hat p}^{(k)}\equiv\frac{2\pi\hbar}{L}\sum_{P}|k;P)\!)(\!(k;P|,
	\label{def:projection-test}
	\end{equation}
and the collision operator $\psi^{(k)}$ is given by (\ref{eq:collop-matel}).
This is equivalent to the Boltzmann equation for the perfect quantum Lorentz gas \cite{Zhang95, Esposito99}, for which the first term in the square bracket in (\ref{eq:collop-matel}) is called the flow term, and the second term is called Boltzmann's collision term.

\subsection{Eigenstates of the collision operator}

Let us denote the right- and left-eigenstates of the collision operator (\ref{eq:nmatbcoll}) as $\dket{\phi_{\alpha}}$ and $\dbra{{\tilde\phi}_{\alpha}}$, respectively, i.e.,
	\begin{equation}
	\nbmat{\ba}\dket{\phi_{\alpha}}={\bar z}_{\alpha}\dket{\phi_{\alpha}},\hspace{10pt}\dbra{{\tilde\phi}_{\alpha}}\nbmat{a}={\bar z}_{\alpha}\dbra{{\tilde\phi}_{\alpha}},
	\end{equation}
where the double bra-ket vectors stand for vectors in the Liouville space (see Eq.(\ref{def:wignerrepresentation})).
The characteristic equation for the collision operator is given by
	\begin{equation}
	{\rm det}\bigr[\nbmat{\ba}-\bz_{\alpha}\bigr]=(\bz_{\alpha}+i)^{2}-(\ba^{2}-1)=0.
	\label{eq:characteristic}
	\end{equation}
Then, we have
	\begin{equation}
	\bz_{\pm}=-i\pm(\ba^{2}-1)^{1/2},
	\label{eq:eigenvalues}
	\end{equation}
	\begin{subequations}
	\begin{equation}
	\dket{\phi_{\pm}(a)}=
	\begin{pmatrix}
	\biggr[1\pm\frac{(\ba^{2}-1)^{1/2}}{\ba}\biggr]^{1/2} \\
	\frac{|a|}{a}i\biggr[1\mp\frac{(\ba^{2}-1)^{1/2}}{\ba}\biggr]^{1/2}
	\end{pmatrix},
	\end{equation}
and
	\begin{equation}
	\dbra{{\tilde\phi}_{\pm}(a)}=\Biggr(\frac{|a|}{a}\biggr[1\pm\frac{(\ba^{2}-1)^{1/2}}{\ba}\biggr]^{1/2},i\biggr[1\mp\frac{(\ba^{2}-1)^{1/2}}{\ba}\biggr]^{1/2}\Biggr),
	\end{equation}
	\label{eq:eigenvectors}
	\end{subequations}
where we have explicitly indicated the parameter $a$ in the eigenstates.
Note that we have not normalized the eigenvectors (\ref{eq:eigenvectors}) considering the fact that they cannot be normalized for $a=\pm1$.
Indeed, the inner products of these right- and left-eigenstates
	\begin{equation}
	\dbraket{{\tilde\phi}_{\pm}(a)}{\phi_{\pm}(a)}=\pm\frac{2(\ba^{2}-1)^{1/2}}{|\ba|}=\frac{z_{\pm}-z_{\mp}}{|a|},
	\label{eq:innerprod}
	\end{equation}
vanish at these points.
In (\ref{eq:eigenvalues}), each of the two eigenvalues is associated with one of the two values taken by the square root function, and the assignment is fixed with the following definition,
\begin{equation}
(a^{2}-1)^{1/2}\equiv\begin{cases}
\sqrt{a^{2}-1},\hspace{13pt}(|a|\geq1)\\
i\sqrt{1-a^{2}}.\hspace{10pt}(|a|\leq1)
\end{cases}
\end{equation}
We also impose the following condition to the relative phase of the components of the vectors (\ref{eq:eigenvectors}),
\begin{equation}
\label{eq:relphase}
\frac{\biggr[1\pm\frac{(\ba^{2}-1)^{1/2}}{\ba}\biggr]^{1/2}}{\biggr[1\mp\frac{(\ba^{2}-1)^{1/2}}{\ba}\biggr]^{1/2}}=|a|\Biggr[1\pm\frac{(a^{2}-1)^{1/2}}{a}\Biggr],
\end{equation}
since the expression has ambiguity because the values taken by the two-valued square root functions are not fixed. 
With the additional condition (\ref{eq:relphase}) each eigenvector is determined up to an overall sign.

Except for these points, the eigenstates (\ref{eq:eigenvectors}) are normalizable as
	\begin{subequations}
	\begin{equation}
	\dket{\chi_{\pm}(a)}\equiv\dket{\phi_{\pm}(a)}\bigr/\bigr[\dbraket{{\tilde\phi}_{\pm}(a)}{\phi_{\pm}(a)}\bigr]^{1/2},
	\end{equation}
	\begin{equation}
	\dbra{{\tilde\chi}_{\pm}(a)}\equiv\dbra{{\tilde\phi}_{\pm}(a)}\bigr/\bigr[\dbraket{{\tilde\phi}_{\pm}(a)}{\phi_{\pm}(a)}\bigr]^{1/2}.
	\end{equation}
	\label{eq:normal-eigenstates}
	\end{subequations}
Then, they satisfy the following bi-orthonormality and bi-completeness relations for $a\not=\pm1$,
	\begin{equation}
	\dbraket{{\tilde\chi}_{\alpha}(a)}{\chi_{\alpha'}(a)}=\delta_{\alpha;\alpha'},
	\end{equation}
	\begin{equation}
	\sum_{\alpha}\dket{\chi_{\alpha}(a)}\!\dbra{{\tilde\chi}_{\alpha}(a)}={\hat I},
	\label{eq:complete}
	\end{equation}
where $\alpha$ and $\alpha'$ take the values ``$+$'' or ``$-$''.

\begin{figure}[t]
	\begin{center}
	\includegraphics[width=0.9 \linewidth]{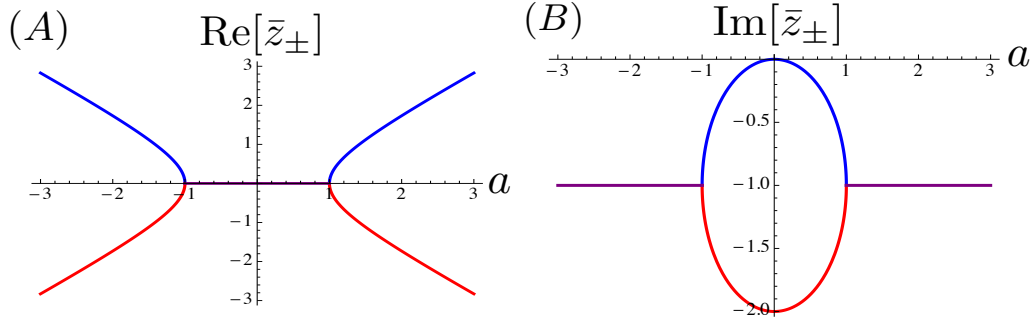}
	\end{center}
	\vspace{-2em}
	\caption{Spectrum of the Boltzmann collision operator. (A) is the Imaginary part of the spectrum and (B) is its real part. In each figure, the blue line represents eigenvalue $\bz_{+}$ and the red line represents eigenvalue $\bz_{-}$. The purple lines represent that these two lines are overlapping.}
	\label{fig:spectrum}
	\end{figure}

We show the $a$-dependence of the real part and the imaginary part of the eigenvalues in Fig.\ref{fig:spectrum}.
In the figures, the blue and the red lines represent the eigenvalues $\bz_{+}$ and $\bz_{-}$, respectively, and the purple lines represent that these two lines are overlapping.

As a function of the parameter $a$, the eigenvalues (\ref{eq:eigenvalues}) have two exceptional points at
	\begin{equation}
	\ba=\pm1.
	\label{eq:degeneracy}
	\end{equation}
At these points, both eigenvalues and eigenstates degenerate as
	\begin{equation}
	\bz_{\pm}=\bz_{0}\equiv-i,
	\end{equation}
	\begin{subequations}
	\begin{equation}
	\dket{\phi^{(a=\pm1)}}\equiv\dket{\phi_{\pm}(a=\pm1)}=
	\begin{pmatrix}
	1 \\
	\pm i
	\end{pmatrix},
	\end{equation}
and
	\begin{equation}
	\dbra{{\tilde\phi}^{(a=\pm1)}}\equiv\dbra{{\tilde\phi}_{\pm}(a=\pm1)}=(\pm1,i).
	\end{equation}
	\label{eq:eigenstates-eq-1}
	\end{subequations}

Since there is only one eigenstate at each degeneracy point (\ref{eq:degeneracy}), the collision operator (\ref{eq:nmatbcoll}) is non-diagonalizable at each point.
Instead, the collision operator is represented by Jordan normal form by introducing the pseudo-eigenstates $\dket{\phi'_{\pm1}}$ and $\dbra{{\tilde\phi}'_{\pm1}}$ through the Jordan chain relations (\ref{eq:general-Jordanchain-right}) and (\ref{eq:general-Jordanchain-left}).
Following the result (\ref{eq:general-pseudo}) in Sec.3, we have
	\begin{equation}
	\dket{\phi'^{(\pm1)}}=\frac{1}{2}
	\begin{pmatrix}
	\pm1 \\
	-i
	\end{pmatrix},\hspace{10pt}
	\dbra{{\tilde\phi}'^{(\pm1)}}=\frac{1}{2}(1,\mp i).
	\label{eq:pseudo-eigenstate}
	\end{equation}
These pseudo-eigenstates satisfy the following bi-orthonomality relations
	\begin{equation}
	\dbraket{{\tilde\phi}^{(\pm1)}}{\phi'^{(\pm1)}}=1,\hspace{5pt}\dbraket{{\tilde\phi}'^{(\pm1)}}{\phi^{(\pm1)}}=1,\hspace{5pt}\dbraket{{\tilde\phi}^{(\pm1)}}{\phi^{(\pm1)}}=0,\hspace{5pt}\dbraket{{\tilde\phi}'^{(\pm1)}}{\phi'^{(\pm1)}}=0,
	\end{equation}
and the bi-completeness relation
	\begin{equation}
	\dket{\phi^{(\pm1)}}\!\dbra{{\tilde\phi}'^{(\pm1)}}+\dket{\phi'^{(\pm1)}}\!\dbra{{\tilde\phi}^{(\pm1)}}={\hat I}.
	\end{equation}
Making use of these pseudo eigenstates, the collision operator is represented by the Jordan normal form matrix as
	\begin{equation}
	\nbmat{\pm1}=
	\bordermatrix{
	& \dket{\phi^{(\pm1)}} & \dket{\phi'^{(\pm1)}} \cr
	\dbra{{\tilde\phi}'^{(\pm1)}} & \bz_{0} & 1 \cr
	\dbra{{\tilde\phi}^{(\pm1)}} & 0 & \bz_{0} \cr
	}.
	\label{eq:jordanblock}
	\end{equation}

\subsection{The extended pseudo-eigenstate representation of the collision operator}

By introducing the extended pseudo-eigenstate representation defined in Sec.4, we have a continuous representation of the collision operator near EPs.
For $a\not=\pm1$, we introduce the extended pseudo-eigenstates, denoted by $\dket{\phi_{+}(a)}$ and $\dket{\phi_{-}(a)}$, through the extended Jordan chain relations (\ref{eq:general-extended-Jordanchain-right-1})-(\ref{eq:general-extended-Jordanchain-left}).
Following the results (\ref{eq:general-extended-pseudo}), we have
	\begin{subequations}
	\begin{equation}
	\dket{\phi'_{\mp}(a)}=\frac{1}{2}
	\begin{pmatrix}
	\frac{|a|}{a}\biggr[1\mp\frac{(\ba^{2}-1)^{1/2}}{\ba}\biggr]^{1/2} \\
	-i\biggr[1\pm\frac{(\ba^{2}-1)^{1/2}}{\ba}\biggr]^{1/2} \\
	\end{pmatrix},
	\end{equation}
and
	\begin{equation}
	\dbra{{\tilde\phi}'_{\pm}(a)}=\frac{1}{2}\Biggr(\biggr[1\pm\frac{(\ba^{2}-1)^{1/2}}{\ba}\biggr]^{1/2},-i\frac{|a|}{a}\biggr[1\mp\frac{(\ba^{2}-1)^{1/2}}{\ba}\biggr]^{1/2}\Biggr).
	\end{equation}
	\label{eq:extended-pseudo-eigenstates}
	\end{subequations}

The two sets of vectors $\{\dket{\phi_{+}},\dket{\phi'_{-}},\dbra{{\tilde\phi}'_{+}},\dbra{{\tilde\phi}_{-}}\}$ and $\{\dket{\phi_{-}},\dket{\phi'_{+}},\dbra{{\tilde\phi}'_{-}},\dbra{{\tilde\phi}_{+}}\}$, respectively, satisfy the following bi-orthonormality and bi-completeness relations,
	\begin{equation}
	\dbraket{{\tilde\phi}_{\pm}(a)}{\phi'_{\pm}(a)}=1,\hspace{5pt}\dbraket{{\tilde\phi}'_{\pm}(a)}{\phi_{\pm}(a)}=1,\hspace{5pt}\dbraket{{\tilde\phi}_{\mp}(\ba)}{\phi_{\pm}(\ba)}=0,\hspace{5pt}\dbraket{{\tilde\phi}'_{\pm}(\ba)}{\phi'_{\mp}(\ba)}=0,
	\label{eq:orthogonal-extended-jordan-basis}
	\end{equation}
and
	\begin{equation}
	\dket{\phi_{\pm}}\!\dbra{{\tilde\phi}'_{\pm}}+\dket{\phi'_{\mp}}\!\dbra{{\tilde\phi}_{\mp}}={\hat I}.
	\label{eq:extended-pseudo-complete}
	\end{equation}
In terms of either basis set, the collision operator is represented by the Jordan block-like matrix for arbitrary values of $a$ as
\begin{equation}
\nbmat{\ba}=
\bordermatrix{
	& \dket{\phi_{\pm}} & \dket{\phi'_{\mp}} \cr
	\dbra{{\tilde\phi}'_{\pm}} & \bz_{\pm} & 1 \cr
	\dbra{{\tilde\phi}_{\mp}} & 0 & \bz_{\mp} \cr
	}.
\label{eq:extended-jordanblock}
\end{equation}
By taking the limit $a\rightarrow1$ or $a\rightarrow-1$ for (\ref{eq:extended-jordanblock}), we recover the Jordan block representation (\ref{eq:jordanblock}) just at the EPs.
The extended pseudo-eigenstates (\ref{eq:extended-pseudo-eigenstates}) also reduce to the usual pseudo-eigenstates (\ref{eq:pseudo-eigenstate}) in this limit.

\subsection{Two different descriptions of time evolution in terms of the eigenstate representation and the extended pseudo-eigenstate representation}

In the previous sections, we obtained two different representations of the collision operator:
one is the eigenstate representation which is not normalizable at the exceptional points $a=\pm1$,
the other is the extended pseudo-eigenstate representation that is normalizable at the exceptional points.
As one might expect, the usual representation in terms of the eigenstates leads to a serious difficulty in the vicinity of the exceptional points, while we have no such difficulty with the normalizable representation in terms of the extended pseudo-eigenstates.

Let us evaluate the time evolution of the solution of the Boltzmann equation (\ref{eq:kineticeq}), which is written as
\begin{equation}
{\hat p}^{(k)}\dket{f(t)}=e^{-i\psi^{(k)}t}{\hat p}^{(k)}\dket{f(0)}=e^{-i{\bar \psi}^{(a)}{\bar t}}{\hat p}^{(k)}\dket{f(0)},
\end{equation}
where ${\bar t}\equiv(\gamma_{P}/2)t$, and the projection operator ${\hat p}^{(k)}$ is defined in (\ref{def:projection-test}).

First, we consider the traditional eigenstate expansion.
Using the $\{\dket{\phi_{+}},\dket{\phi_{-}},\dbra{{\tilde\phi}_{+}},\dbra{{\tilde\phi}_{-}}\}$ representation, we have
\begin{equation}
{\hat p}^{(k)}\dket{f(\bt)}=
\begin{pmatrix}
\frac{|a|}{\bz_{+}-\bz_{-}}e^{-i\bz_{+}\bt} & 0\\
0 & \frac{|a|}{\bz_{-}-\bz_{+}}e^{-i\bz_{-}\bt}
\end{pmatrix}
\begin{pmatrix}
f_{+}(0)\\
f_{-}(0)
\end{pmatrix},
\label{eq:62}
\end{equation}
with
\begin{equation}
f_{\pm}(t)\equiv\dbraket{{\tilde\phi}_{\pm}}{f(t)},
\end{equation}
i.e.,
\begin{equation}
{\hat p}^{(k)}\dket{f(t)}=\frac{|a|}{\bz_{+}-\bz_{-}}e^{-i\bz_{+}\bt}\dket{\phi_{+}}\!\dbraket{{\tilde\phi}_{+}}{f(0)}+\frac{|a|}{\bz_{-}-\bz_{+}}e^{-i\bz_{-}\bt}\dket{\phi_{-}}\!\dbraket{{\tilde\phi}_{-}}{f(0)}.
\label{eq:64}
\end{equation}

On the other hand, using the extended pseudo-eigenstates with the set $\{\dket{\phi_{+}},\dket{\phi'_{-}},\dbra{{\tilde\phi}'_{+}},\dbra{{\tilde\phi}_{-}}\}$, we have
\begin{equation}
{\hat p}^{(k)}\dket{f(\bt)}=
\begin{pmatrix}
e^{-i\bz_{+}\bt} & F(\bt)\\
0 & e^{-i\bz_{-}\bt}
\end{pmatrix}
\begin{pmatrix}
f'_{+}(0)\\
f_{-}(0)
\end{pmatrix},
\label{eq:65}
\end{equation}
where
\begin{equation}
F(\bt)=\frac{e^{-i\bz_{+}\bt}-e^{-i\bz_{-}\bt}}{\bz_{+}-\bz_{-}},
\label{eq:timedev-offdiagonal}
\end{equation}
with
\begin{equation}
f'_{+}(t)\equiv\dbraket{{\tilde\phi}'_{+}}{f(t)},
\end{equation}
i.e.,
\begin{equation}
{\hat p}^{(k)}\dket{f(t)}=e^{-i\bz_{+}\bt}\dket{\phi_{+}}\!\dbraket{{\tilde\phi}'_{+}}{f(0)}+e^{-i\bz_{-}\bt}\dket{\phi'_{-}}\!\dbraket{{\tilde\phi}_{-}}{f(0)}+F(\bt)\dket{\phi_{+}}\!\dbraket{{\tilde\phi}_{-}}{f(0)}.
\label{eq:68}
\end{equation}
Comparing Eq.(\ref{eq:eigenvectors}b) with Eq.(\ref{eq:extended-pseudo-eigenstates}b), we have the relation
\begin{equation}
f'_{+}(t)=\frac{|a|}{\bz_{+}-\bz_{-}}f_{+}(t)+\frac{1}{\bz_{-}-\bz_{+}}f_{-}(t).
\end{equation}

Introducing the Wigner representation of $\dket{f(t)}$ defined by
\begin{equation}
f_{k;P}(\bt)\equiv\dbraket{k;P}{f(\bt)},
\label{eq:fourier-comp-red}
\end{equation}
both relations (\ref{eq:62}) and (\ref{eq:65}) give us the same function,
\begin{equation}
f_{k;P}(\bt)=\frac{1}{2}\Bigr[e^{-i\bz_{+}\bt}+e^{-i\bz_{-}\bt}+2aF(\bt)\Bigr]f_{k;P}(0)+iF(\bt)f_{k;-P}(0).
\label{eq:time-development-notEP}
\end{equation}
This leads to
\begin{equation}
f_{k;P}(\bt)=\Bigr[e^{-\bt}\mp i\bt e^{-\bt}\Bigr]f_{k,P}(0)+\bt e^{-\bt}f_{k,-P}(0),
\label{eq:time-development-EP}
\end{equation}
at the EPs with $a=\pm1$.
We note that Eq.(\ref{eq:fourier-comp-red}) is a Fourier component of the Wigner function (see Eqs.(\ref{eq:10}) and (\ref{def:wignerrepresentation})).

Let us now compare the expressions in Eqs.(\ref{eq:62}) and (\ref{eq:65}).
Each matrix element in the eigenstate expansion (\ref{eq:62}) diverges at the EPs $\bz_{+}=\bz_{-}$.
Hence, the expression (\ref{eq:62}) following from the eigenstate expansion generally leads to serious difficulty when we consider in the vicinity of the EPs.
On the other hand, we have no such difficulty with the expression (\ref{eq:65}) since each matrix element is well defined even at the EPs.

\subsection{Numerical calculation}

\begin{figure}[t]
\begin{center}
\includegraphics[width=1 \linewidth]{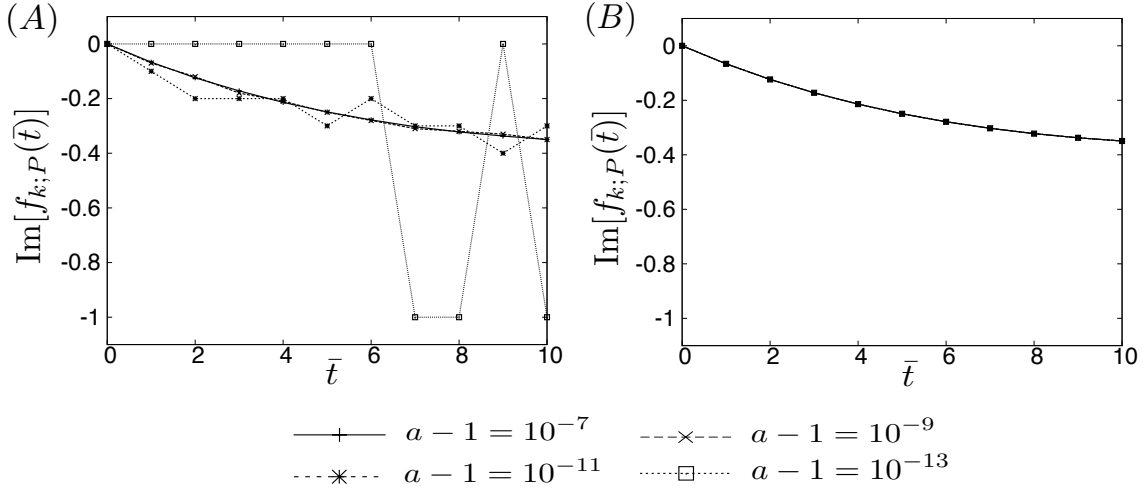}
\end{center}
\vspace{-1em}
\caption{Comparison of the numerical result of ${\rm Im}[f_{k;P}({\bar t})]$ calculated by (A) the eigenstate representation with (B) the extended pseudo-eigenstate representation.
The calculations are performed with a precision of eight significant digits.}
\label{fig:error}
\end{figure}

We now demonstrate the relevancy of the normalizable representation in terms of the pseudo-eigenstates by evaluating the time evolution of $f_{k;P}(t)$ with an initial condition $f_{k;P}(0)=1$ and $f_{k;-P}(0)=0$ for the case $1-a=10^{-7}$, $1-a=10^{-9}$, $1-a=10^{-11}$ and $1-a=10^{-13}$, where $a=1$ is an EP with $\bz_{+}=\bz_{-}$ (see \ref{eq:eigenvalues}).
Here the real part of $f_{k;P}(t)$ for the initial condition has no divergence caused by the EP.
Then, here we show the time evolution of its imaginary part Im$[f_{k;P}(t)]$.

In Fig.\ref{fig:error}(A), we present the result using the eigenstate representation (\ref{eq:62}).
We evaluate this by calculating the time evolution of each component separately in this representation, and sum up the results of all components.
As we can see, each component of (\ref{eq:62}) diverges at the EP.
Therefore, numerical behavior with the eigenstate representation near the EP is not reliable when the parameter $a$ approaches the value at the EP.

In Fig.\ref{fig:error}(B), we present the corresponding result in terms of our extended pseudo-eigenstate representation given in Eq.(\ref{eq:65}), where each component has a well-defined value at the EP.
As expected, there is no singularity in the vicinity of the EPs.
Actually the curve in Fig.\ref{fig:error}(B) consists of 4 lines with different values of $a$, but we cannot distinguish them in the resolution drawn in this figure.
Since the original equation does not have any singularity at the EP, the true behavior does not have any singularity as in the case of Fig.\ref{fig:error}(B).

\subsection{Remarks on physical behavior}

Since we have dealt with the Liouvillian system as an example of operator with EPs, here we give some remarks on physical aspect of EPs in the system.
These remarks give some perspectives to the reader to understand physical consequence of EPs in the spectrum of the Liouvillian on the transport processes.

We first discuss spectral properties of Boltzmann's collision operator.
In the original variables $z=(\gamma_{P}/2){\bar z}$ and $k=ak_{P}$, eigenvalues of the collision operator $z_{\pm}$ are written as
	\begin{equation}
	z_{\pm}=-i\frac{\gamma_{P}}{2}\pm|v|(k^{2}-k_{P}^{2})^{1/2},
	\end{equation}
where $|v|\equiv|P|/m$.

If the wavenumber $k$ is small, and in the region $|k|<k_{P}$, both of the eigenvalues take imaginary values.
In this situation, we expand these eigenvalues in power series of $k^{2}$ as
	\begin{equation}
	z_{+}=-i\frac{|v|^{2}}{\gamma_{P}}k^{2}-i\frac{|v|^{4}}{\gamma_{P}^{3}}k^{4}+O\bigpar{\frac{k}{k_{P}}}^{6},
	\end{equation}
	\begin{equation}
	z_{-}=-i\gamma_{P}+i\frac{|v|^{2}}{\gamma_{P}}k^{2}+i\frac{|v|^{4}}{\gamma_{P}^{3}}k^{4}+O\bigpar{\frac{k}{k_{P}}}^{6}.
	\end{equation}
Here, first few terms in the expansion of $z_{+}$ become dominant in hydrodynamic situation, namely $|k|\ll k_{P}$ and $t\gg\gamma_{P}^{-1}$, and they give transport coefficients of hydrodynamic equation \cite{Resibois77}.
For instance, the first term, which is the second order in $k$, gives diffusion coefficient
	\begin{equation}
	D_{P}\equiv\frac{|v|^{2}}{\gamma_{P}}.
	\label{eq:diffusion}
	\end{equation}
Hence, the eigenstate with $z_{+}$ corresponds to diffusive motion of the system in the hydrodynamic situation.
The higher contribution in the series expansion corresponds to Burnett terms \cite{Resibois77}.
Another eigenvalue $z_{-}$ corresponds to a decaying mode, which is damped in the hydrodynamic situation.

On the other hand, if the wavenumber becomes larger, and lies in the region $|k|>k_{P}$, both of the eigenvalues have real parts
	\begin{equation}
	z_{\pm}=-i\frac{\gamma_{P}}{2}\pm|kv|\Biggr[1-\biggr(\frac{k_{P}}{k}\biggr)^{2}\Biggr]^{1/2},
	\end{equation}
Hence, both eigenstates are damped oscillating modes with damping rate $\gamma_{P}/2$.
In the real space, both of them correspond to damping wave propagation with a phase velocity
	\begin{equation}
	v_{{\rm phase}}=\pm v\Biggr[1-\biggr(\frac{k_{P}}{k}\biggr)^{2}\Biggr]^{1/2}.
	\end{equation}
As the wavenumber $k$ becomes larger, the phase velocity $v_{{\rm phase}}$ approaches the velocity of the free particle.
Hence, $k=k_{P}$ leads to a phase transition from the over damping situation to the oscillating damping situation.

As a result of above arguments on spectral properties, we have seen that the eigenmodes for $|k|<k_{P}$ correspond to diffusive process and the eigenmodes for $|k|>k_{P}$ correspond to damping wave propagation with damping rate $\gamma_{P}/2$.
Here, the EPs play a role of boundaries of these qualitatively different processes.
The coexistence of these different processes is a generic property of the systems that have EPs in the wavenumber space of the spectrum of the Liouvillian, since these EPs always appear as boundaries of the pure imaginary values and complex values of the eigenvalues.

We now show that the existence of EPs leads to the telegraph equation, which describes combined processes of the diffusive process and the damping wave propagation.
First we introduce the Wigner distribution function for the test particle
	\begin{equation}
	f^{W}(X,P,t)\equiv\int^{\infty}_{-\infty}dke^{ikX}f_{k;P}(t),
	\end{equation}
where $f_{k;P}(t)$ is defined by (\ref{eq:fourier-comp-red}).
In terms of the original variables $z=(\gamma_{P}/2){\bar z}$ and $k+ak_{P}$, the characteristic equation (\ref{eq:characteristic}) is written as
	\begin{equation}
	z^{2}-i\gamma_{P}z-|v|^{2}k^{2}=0.
	\label{eq:B1}
	\end{equation}
The inverse Fourier-Laplace transformation of (\ref{eq:B1}) leads to the telegraph equation
	\begin{equation}
	\frac{\partial^{2}}{\partial t^{2}}f^{W}(X,P,t)+\gamma_{P}\frac{\partial}{\partial t}f^{W}(X,P,t)=|v|^{2}\frac{\partial^{2}}{\partial X^{2}}f^{W}(X,P,t).
	\end{equation}
In other words, Eq.(\ref{eq:B1}) is the same as the characteristic equation of the telegraph equation with regard to the $X$- and $t$-dependence as $\exp[i(kX-zt)]$.
Hence our Boltzmann equation (\ref{eq:kineticeq}) is equivalent to the telegraph equation \cite{Zhang95,Morse53} with regard to the dependence of the Wigner function on $X$ and $t$.

The equivalence of Boltzmann's equation and the telegraph equation in their time development in real space is remarkable, since the telegraph equation represents a prototypical behavior of the system with a second order EP in the spectrum of the Liouvillian with respect to the wavenumber.
This is because, as we have shown in Sec.2, when the spectrum of the Liouvillian has a second order EP in the wavenumber space, the characteristic equation is always reduced to the quadratic form (\ref{eq:B1}) locally near the EP.

We now show that the telegraph equation reduces to the diffusion equation in long-time behavior.
To see this we observe Fig.1B.
All decaying modes except for eigenstates with small pure imaginary eigenvalues have vanished in a long time region.
For the remaining modes, $|z|\ll\gamma_{P}$, and the first term of (\ref{eq:B1}) is much smaller than the second term.
Hence, the characteristic equation (\ref{eq:B1}) reduces to
\begin{equation}
i\gamma_{P}z+|v|^{2}k^{2}=0.
\label{eq:B2}
\end{equation}
Then, inverse Fourier-Laplace transformation of (\ref{eq:B2}) leads to the diffusion equation,
\begin{equation}
\frac{\partial}{\partial t}f^{W}(X,P,t)=D_{P}\frac{\partial^{2}}{\partial X^{2}}f^{W}(X,P,t),
\end{equation}
where $D_{P}$ is the diffusion coefficient given by (\ref{eq:diffusion}).

Let us now summarize this subsection.
When the system under consideration has the second order EPs in the wavenumber space of the spectrum of the Liouvillian, the time development of the system in the real space is described by the telegraph equation that combines the diffusive process and the damping wave propagation.
Then, in the long time behavior such as $t\gg\gamma_{P}^{-1}$, the telegraph equation asymptotically reduces to the diffusion equation.
Since the second order EPs are the most likely case and the characteristic equation of the operator is always reduced to the quadratic equation of the form (\ref{eq:B1}) near the second order EPs, above description of the time development of the Liouvillian system with EPs is quite generic.

\section{Summary and Discussion}

We have introduced a non-divergent representation of non-Hermitian operators that remains finite even at the EPs in a parameter space.
The representation has been obtained by extending the pseudo-eigenstates to the entire parameter space.

We have applied this representation to the collision operator of the Boltzmann equation for the 1D perfect quantum Lorentz gas.
Then we have shown that this representation removes the difficulty resulting from the divergence of the normalization constant at the EPs in the usual eigenstate expansion.
Indeed, we have demonstrated a dramatic improvement in the accuracy of the numerical evaluation of the time evolution of the distribution function in terms of our representation.

There we have also shown physical aspects of the EPs in the Liouvillian system.
As we have shown in the subsection 5.8. in Sec.5, when the Liovillian system has the second order EPs in the wavenumber space, such a spectral structure leads to the telegraph equation that combines two qualitatively different processes in time development of the system, namely the diffusive process and the damping wave propagation.
In the long time behavior, the damping wave vanishes and the time development is reduced to usual diffusion.

In our recent studies, we have realized that wide range of Liouvillian systems such as the two-dimensional classical perfect Lorentz gas \cite{Zhang95} or one-dimensional polaron system also have the second order EPs in the wavenumber space.
Therefore, appearance of the second order EPs in the wavenumber space is a quite common nature of the Liouvillian system.
Although these systems are rather complicated than the 1D perfect quantum Lorentz gas, we expect that our essential view on the role of EPs on the transport process can be applicable to these systems.
We will discuss it in a separate publication.

In recent years, multiple coalescence where an arbitrary number $N$ of eigenstates coalesce at a single EP, called EP$N$, has also been studied \cite{Heiss08}.
In appendix A we also show that our extended pseudo-eigenstate representation can be generalized to the EP$N$s.

The EPs are non-Hermitian degeneracy points.
It is well-known that degeneracy in a Hermitian operator leads to the Berry phase effect \cite{Berry84} as reported in many experiments, see e.g. \cite{Berry-review}.
Recently, it is clarified that the Berry phase-like effect plays an important role in the study of quantum pumping processes based on quantum kinetic equations (quantum master equations) \cite{Sinitsyn09, Ren10, yuge12, yuge13, Watanabe}.
Similar to the degenerate Hermitian operator, it is interesting to investigate the effects coming from the degeneracy in non-Hermitian operators and the phase change of the eigenstates around the EPs.
Indeed, one can find many theoretical \cite{Heiss90, Heiss99, Heiss01, Savannah12, Berry04, Heiss04, Cartarius07, Rubinstein07, Heiss08, Heiss10, Lefebvre09, Cartarius11, Demange12, Gilary13} and experimental papers \cite{Dembowski01, Dembowski03, Dietz07, Lee09, Schindler11, Peng14} on this subject.
We will discuss the implication of our extended pseudo-eigenstate representation on the phase of the eigenstates in the vicinity of the EPs elsewhere.

\section*{Acknowledgement}

We would like to express our sincere gratitude to Prof. S. Tanaka and Prof. N. Hatano for their fruitful discussions and helpful comments on this subject.
We also thank Dr. S. Garmon for her critical reading of the manuscript.
This work was supported by JSPS KAKENHI Grant Number 24540411.
T.P. expresses his sincere gratitude to Yukawa Institute for Theoretical Physics (YITP), Kyoto University for warm hospitality during his stay at YITP, where this study has been launched.

\appendix

\section{Generalization of the Extended Pseudo-eigenstate Representation}

In this appendix, we discuss generalization of the extended pseudo-eigenstate representation to the coalescence of an arbitrary number $N$ of eigenstates at a so-called EP$N$ \cite{Heiss08}.

Let us now consider multiple coalescence.
Suppose a non-Hermitian operator ${\hat M}$ which has an EP where $N$ eigenstates, denoted by ${\bf u}_{1},{\bf u}_{2},\cdots,{\bf u}_{N}$, coalesce.
Here we show that the relations (\ref{eq:general-extended-Jordanchain-right-1})-(\ref{eq:normal-extended-jordan-basis}) and (\ref{eq:general-orthogonal-extended-jordan-basis})-(\ref{eq:general-extend-jordan}) can be generalized to this situation.
For this matrix, the right extended pseudo-eigenstates for eigenvalues $z_{2},z_{3},\cdots,z_{4}$ are defined by the following relations,
\begin{eqnarray}
\begin{split}
&[{\hat M}-z_{2}{\hat I}]{\bf u}'_{2}={\bf u}_{1},\\
&[{\hat M}-z_{3}{\hat I}]{\bf u}'_{3}={\bf u}'_{2},\\
&\hspace{30pt}\vdots\\
&[{\hat M}-z_{N}{\hat I}]{\bf u}'_{N}={\bf u}'_{N-1}.\\
\end{split}
\end{eqnarray}
Similarly, the left extended pseudo-eigenstates for eigenvalues $z_{N-1},z_{N-2},\cdots,z_{1}$ are defined by
\begin{eqnarray}
\begin{split}
&{\tilde{\bf u}}^{\prime\;\dagger}_{N-1}[{\hat M}-z_{N-1}{\hat I}]={\tilde{\bf u}}^{\dagger}_{N},\\
&{\tilde{\bf u}}^{\prime\;\dagger}_{N-2}[{\hat M}-z_{N-2}{\hat I}]={\tilde{\bf u}}^{\prime\;\dagger}_{N-1},\\
&\hspace{40pt}\vdots\\
&{\tilde{\bf u}}^{\prime\;\dagger}_{1}[{\hat M}-z_{1}{\hat I}]={\tilde{\bf u}}^{\prime\;\dagger}_{2},\\
\end{split}
\end{eqnarray}
where ${\tilde{\bf u}}_{N}$ is the left eigenstate of ${\hat M}$ for the eigenvalue $z_{N}$.
Imposing the following normalization condition to the extended pseudo-eigenstates,
\begin{eqnarray}
\begin{split}
&({\tilde{\bf u}}'_{1},{\bf u}_{1})=1,\\
&({\tilde{\bf u}}'_{2},{\bf u}'_{2})=1,\\
&\hspace{20pt}\vdots\\
&({\tilde{\bf u}}_{N},{\bf u}'_{N})=1,\\
\end{split}
\end{eqnarray}
where $({\tilde{\bf v}}_{1},{\bf v}_{2})$ is the inner product of vectors ${\tilde{\bf v}}_{1}$ and ${\bf v}_{2}$,
we have the following orthogonal relations
\begin{equation}
({\tilde{\bf u}}'_{j},{\bf u}_{1})=0,\hspace{5pt}({\tilde{\bf u}}_{N},{\bf u}'_{j'})=0,\hspace{5pt}({\tilde{\bf u}}'_{j},{\bf u}'_{j'})=0,
\end{equation}
where $j\not=1$, $j'\not=N$ and $j\not=j'$.
These vectors form a complete basis,
\begin{equation}
{\bf u}_{1}{\tilde{\bf u}}^{\prime\;\dagger}_{1}+{\bf u}'_{2}{\tilde{\bf u}}^{\prime\;\dagger}_{2}+\cdots+{\bf u}'_{j}{\tilde{\bf u}}^{\prime\;\dagger}_{j}+\cdots+{\bf u}'_{N}{\tilde{\bf u}}^{\dagger}_{N}={\hat I}.
\end{equation}
Using the basis, the matrix ${\hat M}$ is represented as
\begin{equation}
{\hat M}=
\bordermatrix{
	& {\bf u}_{1} & {\bf u}'_{2} & {\bf u}'_{3} & \cdots & {\bf u}'_{N-1} & {\bf u}'_{N} \cr
	{\tilde{\bf u}}^{\prime\;\dagger}_{1} & z_{1} & 1 & 0 & \cdots & 0 & 0 \cr
	{\tilde{\bf u}}^{\prime\;\dagger}_{2} & 0 & z_{2} & 1 & \cdots & 0 & 0 \cr
	{\tilde{\bf u}}^{\prime\;\dagger}_{3} & 0 & 0 & z_{3} & \cdots & 0 & 0 \cr
	\vdots & \vdots  & \vdots & \vdots  & \ddots & \vdots & \vdots \cr
	{\tilde{\bf u}}^{\prime\;\dagger}_{N-1} & 0 & 0 & 0 & \cdots & z_{N-1} & 1 \cr
	{\tilde{\bf u}}^{\dagger}_{N} & 0 & 0 & 0 & \cdots & 0 & z_{N} \cr
}.
\end{equation}
This representation reduces to the $N$-th order Jordan normal form at the EP.

\section{General Formalism of the Complex Spectral Representation of the Liouvillian}

We consider a quantum system described by a Hamiltonian $H$.
The time evolution of the system is governed by the Liouville-von Neumann equation for the density matrix $\rho(t)$,
\begin{equation}
i\frac{\partial}{\partial t}\rho(t)=L_{H}\rho(t).
\label{eq:liouville-eq}
\end{equation}
Here $L_{H}$ is the Liouville-von Neumann operator (Liouvillan in short) which is defined by the commutation relation with the Hamiltonian of the system $H$,
\begin{equation}
L_{H}\rho\equiv\frac{1}{\hbar}[H,\rho].
\end{equation}

In the Liouville space, the eigenvalue problem of the Liouvillian for each correlation subspace \cite{petpri97}, which is specified by the index $\nu$, is given by
\begin{equation}
L_{H}\dket{F^{(\nu)}_{\alpha}}=Z^{(\nu)}_{\alpha}\dket{F^{(\nu)}_{\alpha}},\hspace{10pt}\dbra{{\tilde F}^{(\nu)}_{\alpha}}L_{H}=\dbra{{\tilde F}^{(\nu)}_{\alpha}}Z^{(\nu)}_{\alpha},
\label{eq:eigenvalue-eq}
\end{equation}
where the double bra-ket vectors stand for vectors in the Liouville space and the index $\alpha$ specifies the eigenstate in the correlation subspace denoted by $\nu$.
We denote the right-eigenstate as $\dket{F^{(\nu)}_{\alpha}}$ and the left-eigenstate as $\dbra{{\tilde F}^{(\nu)}_{\alpha}}$.
We solve the eigenvalue problem by using the well-known Brillouin-Wigner-Feshbach formalism \cite{petpri97} with projection operators $P^{(\nu)}$ and $Q^{(\nu)}$ which satisfy
\begin{equation}
P^{(\nu)}+Q^{(\nu)}=1.
\end{equation}
By applying these projection operators on (\ref{eq:eigenvalue-eq}), the eigenvalue equation of the Liouvillian takes the form
\begin{equation}
\Psi^{(\nu)}(Z^{(\nu)}_{\alpha})P^{(\nu)}\dket{F^{(\nu)}_{\alpha}}=Z^{(\nu)}_{\alpha}P^{(\nu)}\dket{F^{(\nu)}_{\alpha}},
\label{eq:dispersion}
\end{equation}
where
\begin{eqnarray}
\Psi^{(\nu)}(z)\!\equiv\!P^{(\nu)}L_{H}P^{(\nu)}+P^{(\nu)}L_{H}Q^{(\nu)}\frac{1}{z-Q^{(\nu)}L_{H}Q^{(\nu)}}Q^{(\nu)}L_{H}P^{(\nu)}
\label{eq:collop}
\end{eqnarray}
is the effective Liouvillian and the second term is the self-frequency part.
In the eigenvalue problem of the Hamiltonian, a similar expression to (\ref{eq:collop}) is called the effective Hamiltonian, and the second term in the case is called the self-energy operator \cite{Hatano13}.
The effective Liouvillian is also called the {\it collision operator} which is of central importance in the kinetic theory in non-equilibrium statistical mechanics \cite{petpri97}.
One can see from its eigenvalue equation (\ref{eq:dispersion}) that the collision operator has the same eigenvalues as those of the Liouvillian.
Moreover, the eigenvalue equation is non-linear since the collision operator itself depends on the eigenvalue.

It is well-known for an unstable quantum system with a continuous spectrum that the effective Hamiltonian becomes a non-Hermitian operator due to the resonance singularity in the self-energy.
Similarly, the collision operator also becomes a non-Hermitian operator in the Liouville space in the thermodynamic limit.
As a result, the collision operator has eigenstates with complex eigenvalues, which are called {\it resonance states}.
For the collision operator, the imaginary part of the complex eigenvalue gives a transport coefficient of the system.

In terms of the right- and left-eigenstates of the collision operator $\Psi^{(\nu)}(z)$, the eigenstates of the Liouvillian $L_{H}$ are, respectively, expressed by
\begin{equation}
\dket{F^{(\nu)}_{\alpha}}=\bigr[P^{(\nu)}+{\cal C}^{(\nu)}(Z^{(\nu)}_{\alpha})\bigr]P^{(\nu)}\dket{F^{(\nu)}_{\alpha}},\hspace{10pt}\dbra{{\tilde F}^{(\nu)}_{\alpha}}=\dbra{{\tilde F}^{(\nu)}_{\alpha}}P^{(\nu)}\bigr[P^{(\nu)}+{\cal D}^{(\nu)}(Z^{(\nu)}_{\alpha})\bigr],
\end{equation}
with the {\it creation of correlation operator}
\begin{subequations}
\begin{equation}
{\cal C}^{(\nu)}(z)=\frac{1}{z-Q^{(\nu)}L_{H}Q^{(\nu)}}Q^{(\nu)}L_{H}P^{(\nu)},
\end{equation}
and the {\it destruction of correlation operator}
\begin{equation}
{\cal D}^{(\nu)}(z)=P^{(\nu)}L_{H}Q^{(\nu)}\frac{1}{z-Q^{(\nu)}L_{H}Q^{(\nu)}},
\end{equation}
\end{subequations}
which are non-diagonal transitions between the $P^{(\nu)}$ subspace and the $Q^{(\nu)}$ subspace \cite{petpri97}.


\begin{thebibliography}{99}

\bibitem{Balescu75}
R. Balescu, \emph{Equilibrium and Nonequilibrium Statistical Mechanics}, (John Wiley \& Sons Inc., 1975).

\bibitem{Resibois77}
P. Resibois, \emph{Classical Kinetic Theory of Fluids}, (John Wiley \& Sons Inc., 1977).

\bibitem{Heiss90}
W. D. Heiss and A. L. Sannino, J. Phys. A: Math. Gen. {\bf 23}, 1167 (1990).

\bibitem{Heiss91}
W. D. Heiss and W. H. Steeb, J. Math. Phys. {\bf 32}, 3003 (1991).

\bibitem{Heiss99}
W. D. Heiss, Eur. Phys. J. D {\bf 7}, 1 (1999).

\bibitem{Heiss01}
W. D. Heiss and H. L. Harney, Eur. Phys. J. D {\bf 17}, 149 (2001).

\bibitem{Klaiman08}
S. Klaiman, U. Gunther, and N. Moiseyev, Phys. Rev. Lett. {\bf 101}, 080402 (2008).

\bibitem{Velle13}
G. D. Valle and S. Longhi, Phys. Rev. A {\bf 87}, 022119 (2013).

\bibitem{Candanedo14}
O. V\'azquez-Candanedo, J. C. Hern\'andez-Herrej\'on, F. M. Izrailev, and D. N. Christodoulides, Phys. Rev. A {\bf 89}, 013832 (2014).

\bibitem{Peng14}
B. Peng, S. K. Ozdemir, F. L. Lei, F. Monifi, M. Gianfreda, G. L. Long, S. Fan, F. Nori, C. M. Bender, and L. Yang, Nature Phys. {\bf 10}, 394 (2014).

\bibitem{petpri97}
T. Petrosky and I. Prigogine, Adv. Chem. Phys. {\bf 99}, 1 (1997).

\bibitem{Breuer02}
e.g., see H. -P. Breuer, and F. Petruccione, \emph{The Theory of Open Quantum Systems}, (Oxford University Press, Oxford, 2002).

\bibitem{Kato}
T. Kato, \emph{Perturbation Theory of Linear Operators}, (Springer, Berlin, 1966).

\bibitem{Berry04}
M. V. Berry, Czech. J. Phys. {\bf 57}, 1039 (2004).

\bibitem{Savannah12}
S. Garmon, I. Rotter, N. Hatano, and D. Segal, Int. J. Theor. Phys. {\bf 51}, 3536 (2012).

\bibitem{Heiss04}
W. D. Heiss, J. Phys. A: Math. Gen. {\bf 37}, 2455 (2004).

\bibitem{Cartarius07}
H. Cartarius, J. Main, and G. Wunner, Phys. Rev. Lett. {\bf 99}, 173003 (2007).

\bibitem{Rubinstein07}
J. Rubinstein, P. Sternberg, and Q. Ma, Phys. Rev. Lett. {\bf 99}, 167003 (2007).

\bibitem{Heiss08}
W. D. Heiss, J. Phys. A: Math. Theor. {\bf 41}, 244010 (2008).

\bibitem{Heiss10}
W. D. Heiss, Eur. Phys. J. D {\bf 60}, 257 (2010).

\bibitem{Lefebvre09}
R. Lefebvre, O. Atabek, M. Sindelka, and N. Moiseyev, Phys. Rev. Lett. {\bf 103}, 123003 (2009).

\bibitem{Cartarius11}
H. Cartarius and N. Moiseyev, Phys. Rev. A {\bf 84}, 013419 (2011).

\bibitem{Demange12}
G. Demange and E. Graefe, J. Phys. A: Math. Theor. {\bf 45}, 025303 (2012).

\bibitem{Gilary13}
I. Gilary, A. A. Mailybaev, and N. Moiseyev, Phys. Rev. A {\bf 88}, 010102 (2013).

\bibitem{Fuchs14}
J. Fuchs, J. Main, H. Cartarius, and G. Wunner, J. Phys. A: Math. Theor. {\bf 47}, 125304 (2014).

\bibitem{Dembowski01}
C. Dembowski, H. -D. Graf, H. L. Harney, A. Heine, W. D. Heiss, H. Rehfeld, and A. Rihter, Phys. Rev. Lett. {\bf 86}, 787 (2001).

\bibitem{Dembowski03}
C. Dembowski, B. Dietz, H. -D. Graf, H. L. Harney, A. Heine, W. D. Heiss, and A. Rihter, Phys. Rev. Lett. {\bf 90}, 034101 (2003).

\bibitem{Dietz07}
B. Dietz, T. Friedrich, J. Metz, M. Miski-Oglu, A. Richter, F. Schafer, and C. A. Stafford, Phys. Rev. E {\bf 75}, 027201 (2007).

\bibitem{Schindler11}
J. Schindler, A. Li, M. C. Zheng, F. M. Ellis, and T. Kottos, Phys. Rev. A {\bf 84}, 040101 (2011).

\bibitem{Lee09}
S. Lee, J. Yang, S. Moon, S. Lee, J. Shim, S. W. Kim, J, Lee, and K. An, Phys. Rev. Lett. {\bf 103}, 134101 (2009).

\bibitem{Bhamathi}
G. Bhamathi and E. C. G. Sudarshan, Int. J. Mod. Phys. B {\bf 10}, 1531 (1996).

\bibitem{Balescu63}
R. Balescu, \emph{Statistical Mechanics of Charged Particles}, (John Wiley \& Sons Inc., 1963).

\bibitem{Zhang95}
Z. L. Zhang, Doctoral dissertation, The University of Texas at Austin, 1995.

\bibitem{Hatano13}
N. Hatano, Fortschr Phys. {\bf 61}, 238 (2013).

\bibitem{Esposito99}
R. Esposito, M. Pulvirenti, and A. Teta, Commun. Math. Phys. {\bf 204}, 619 (1999).

\bibitem{Morse53}
e.g., see P. M. Morse, and H. Feshbach, \emph{Methods of Theoretical Physics}, (Cambridge University Press, 1953).

\bibitem{Berry84}
M. V. Berry, Proc. R. Soc. Lond. A {\bf 392}, 45 (1984).

\bibitem{Berry-review}
A. Shapere, F. Wilczek, eds., \emph{Geometric Phases in Physics}, (World Scientific, Singapore, 1989).

\bibitem{Sinitsyn09}
e.g., see N. A. Sinitsyn, J. Phys. A: Math. Theor. {\bf 42}, 193001 (2009) and references therein.

\bibitem{Ren10}
J. Ren, P. H\"anggi, and B. Li, Phys. Rev. Lett. {\bf 104}, 170601 (2010).

\bibitem{yuge12}
T. Yuge, T. Sagawa, A. Sugita and, H. Hayakawa, Phys. Rev. B {\bf 86}, 235308 (2012).	

\bibitem{yuge13}
T. Yuge, T. Sagawa, A. Sugita and, H. Hayakawa, J. Stat. Phys. {\bf 153}, 412 (2013).

\bibitem{Watanabe}
K. L. Watanabe and H. Hayakawa, Prog. Theor. Exp. Phys. 2014, 113A01 (2014).


\end{thebibliography}
\end{document}